\documentclass[prd,twocolumn,floatfix,preprintnumbers,showpacs]{revtex4}
\usepackage{graphics}
\usepackage{longtable}
\usepackage{epsfig}

\begin{document}
\title{Can cosmology detect hierarchical neutrino masses?}
\author{Steen Hannestad}
\email{steen@nordita.dk}
\affiliation{Department of Physics, University of Southern Denmark, 
Campusvej 55, DK-5230 Odense M, Denmark \\
and
\\
NORDITA, Blegdamsvej 17, DK-2100 Copenhagen, Denmark}
\date{{\today}}

\begin{abstract}
We have carefully analysed the potential of future Cosmic
Microwave Background (CMB) and Large Scale Structure (LSS) measurements
to probe neutrino masses. We perform a Fisher matrix analysis on
a 9-dimensional cosmological parameter space and find that
data from the Planck CMB experiment combined with the Sloan Digital
Sky Survey (SDSS) can measure a neutrino mass of 0.12 eV at 95\% conf.
This is almost at the level of the 0.06 eV mass suggested by current
neutrino oscillation data. A future galaxy survey with an order of 
magnitude larger survey volume than the SDSS would allow for a
neutrino mass determination of 0.03-0.05 eV (95\% conf.).
\end{abstract}
\pacs{14.60.Pq,95.35.+d,98.80.-k} 
\maketitle


\section{introduction}

The absolute value of neutrino masses are very difficult to measure
experimentally. On the other hand, mass differences between neutrino
mass eigenstates, $(m_1,m_2,m_3)$, 
can be measured in neutrino oscillation experiments.
Observations of atmospheric neutrinos suggest a squared mass 
difference of $\delta m^2 \simeq 3 \times 10^{-3}$ eV$^2$
\cite{Fukuda:2000np,Fornengo:2000sr,Maltoni:2002ni}. While there are still
several viable solutions to the solar neutrino problem the so-called
large mixing angle solution gives by far the best fit with
$\delta m^2 \simeq 5 \times 10^{-5}$ eV$^2$ \cite{sno,Bahcall:2002hv}. 

In the simplest case where neutrino masses are
hierarchical these results suggest that $m_1 \sim 0$, $m_2 \sim 
\delta m_{\rm solar}$, and $m_3 \sim \delta m_{\rm atmospheric}$.
If the hierarchy is inverted 
\cite{Kostelecky:1993dm,Fuller:1995tz,Caldwell:1995vi,Bilenky:1996cb,King:2000ce,He:2002rv}
one instead finds
$m_3 \sim 0$, $m_2 \sim \delta m_{\rm atmospheric}$, and 
$m_1 \sim \delta m_{\rm atmospheric}$.
However, it is also possible that neutrino
masses are degenerate
\cite{Ioannisian:1994nx,Bamert:vc,Mohapatra:1994bg,Minakata:1996vs,Vissani:1997pa,Minakata:1997ja,Ellis:1999my,Casas:1999tp,Casas:1999ac,Ma:1999xq,Adhikari:2000as}, 
$m_1 \sim m_2 \sim m_3 \gg \delta m_{\rm atmospheric}$, 
in which case oscillation experiments are not
useful for determining the absolute mass scale.

Experiments which rely on kinematical effects of the neutrino mass
offer the strongest probe of this overall mass scale. Tritium decay
measurements have been able to put an upper limit on the electron
neutrino mass of 2.2 eV (95\% conf.) \cite{Bonn:tw}.
However, cosmology at present yields an even stronger limit which
is also based on the kinematics of neutrino mass.
Neutrinos decouple at a temperature of 1-2 MeV in the early universe,
shortly before electron-positron annihilation.
Therefore their temperature is lower than the photon temperature
by a factor $(4/11)^{1/3}$. This again means that the total neutrino
number density is related to the photon number density by
\begin{equation}
n_{\nu} = \frac{9}{11} n_\gamma
\end{equation}

Massive neutrinos with masses $m \gg T_0 \sim 2.4 \times 10^{-4}$ eV
are non-relativistic at present and therefore contribute to the
cosmological matter density \cite{Hannestad:1995rs,Dolgov:1997mb,Mangano:2001iu}
\begin{equation}
\Omega_\nu h^2 = \frac{\sum m_\nu}{92.5 \,\, {\rm eV}},
\end{equation}
calculated for a present day photon temperature $T_0 = 2.728$K. Here,
$\sum m_\nu = m_1+m_2+m_3$.
However, because they are so light
these neutrinos free stream on a scale of roughly 
$k \simeq 0.03 m_{\rm eV} \Omega_m^{1/2} \, h \,\, {\rm Mpc}^{-1}$
\cite{dzs,Doroshkevich:tq,Hu:1997mj}. 
Below this scale neutrino perturbations are completely erased and 
therefore the matter power spectrum is suppressed, roughly by
$\Delta P/P \sim -8 \Omega_\nu/\Omega_m$ \cite{Hu:1997mj}.

This power spectrum suppression allows for a determination of the
neutrino mass from measurements of the matter power spectrum on
large scales. This matter spectrum is related to the galaxy correlation
spectrum measured in large scale structure (LSS) surveys via the
bias parameter, $b^2(k) \equiv P_g(k)/P_m(k)$.
Such analyses have been performed several times before
\cite{Croft:1999mm,Fukugita:1999as}, most recently
using data from the 2dF galaxy survey 
\cite{Elgaroy:2002bi,Hannestad:2002xv,Lewis:2002ah}. 
These investigations find mass limits of 1.5-3 eV, depending on
assumptions about the cosmological parameter space.

In a seminal paper it was calculated by Eisenstein, Hu and Tegmark
that future CMB and LSS experiments could push the bound on the
sum of neutrino masses down to about 0.3 eV \cite{Hu:1997mj}.
This calculation was done using a Fisher matrix analysis on a
6-dimensional cosmological parameter space and a fairly crude
approximation of the MAP CMB data.

In the present paper we discuss the prospects for neutrino mass
detection in more detail. First we discuss the Fisher matrix technique
in some detail and then we move on to discuss future data sets from
CMB and LSS. It turns out to be possible to describe hypothetical
new experiments in terms of just a few general parameters.
Furthermore, the issue of parameter degeneracies is discussed in
some detail.

As discussed above, the main neutrino physics mass parameter for 
CMB and LSS is the sum of all neutrino mass eigenstates.
However, for hierarchical neutrino masses this is almost equivalent
to the mass of the heaviest mass eigenstate, $\sum m_\nu \simeq m_3$.
In the remainder of the paper we therefore use $m_\nu$ and
$\sum m_\nu$ interchangeably, except where otherwise stated.


\section{Fisher matrix analysis}

Measuring neutrino masses from cosmological data is quite involved
since for both CMB and LSS the power spectra depend on a plethora
of different parameters in addition to the neutrino mass.
Furthermore, since the CMB and matter power spectra
depend on many different parameters one might
worry that an analysis which is too restricted in parameter space 
could give spuriously strong limits on a given parameter.
Therefore, it is desirable to study possible parameter degeneracies in
a simple way before embarking on a full numerical likelihood analysis.

It is possible to estimate the precision with which the cosmological
model parameters can be extracted from a given hypothetical data set.
The starting point for any parameter extraction is the vector of
data points, $x$. This can be in the form of the raw data, or in
compressed form, typically the power spectrum ($C_l$ for CMB and
$P(k)$ for LSS).

Each data point has contributions from both signal and noise,
$x = x_{\rm signal} + x_{\rm noise}$. If both signal and noise are
Gaussian distributed it is possible to build a likelihood function
from the measured data which has the following form \cite{oh}
\begin{equation}
{\cal L}(\Theta) \propto \exp \left( -\frac{1}{2} x^\dagger 
[C(\Theta)^{-1}] x \right),
\end{equation}
where $\Theta = (\Omega, \Omega_b, H_0, n_s, \tau, \ldots)$ is a vector
describing the given point in model parameter space and 
$C(\Theta) = \langle x x^T \rangle$ 
is the
data covariance matrix.
In the following we shall always work with data in the form of a
set of power spectrum coefficients, $x_i$, which can be either
$C_l$ or $P(k)$.

If the data points are uncorrelated so that the data covariance matrix
is diagonal, the likelihood function can be reduced to
${\cal L} \propto e^{-\chi^2/2}$, where
\begin{equation}
\chi^2 = \sum_{i=1}^{N_{\rm max}} \frac{(x_{i, {\rm obs}}-x_{i,{\rm theory}})^2}
{\sigma(x_i)^2},
\label{eq:chi2}
\end{equation} 
is a $\chi^2$-statistics and $N_{\rm max}$ 
is the number of power spectrum data
points \cite{oh}.

The maximum likelihood is an unbiased estimator, which means that
\begin{equation}
\langle \Theta \rangle = \Theta_0.
\end{equation}
Here $\Theta_0$ indicates the true parameter vector of the underlying
cosmological model and $\langle \Theta \rangle$ is the average estimate
of parameters from maximizing the likelihood function.

The likelihood function should thus peak at $\Theta \simeq \Theta_0$, and
we can expand it to second order around this value.
The first order derivatives are 
zero, and the expression is thus
\begin{widetext}
\begin{equation}
\chi^2  = \chi^2_{\rm min} + \sum_{i,j}(\theta_i-\theta) \left( \sum_{k=1}^{N_{\rm max}}
\frac{1}{\sigma (x_k)^2} \left[\frac{\partial x_k}{\partial \theta_i}
\frac{\partial x_k}{\partial \theta_j} - (x_{k, {\rm obs}}-x_k)
\frac{\partial^2 x_k}{\partial \theta_i \partial \theta_j} \right]\right) (\theta_j-\theta),
\end{equation}
\end{widetext}
where $i,j$ indicate elements in the parameter vector $\Theta$.
The second term in the second derivative can be expected to be very small
because $(x_{k, {\rm obs}}-x_k)$ is in essence just a random measurement error 
which should average out. The remaining term
is usually referred to as the Fisher information matrix
\begin{equation}
F_{ij} = \frac{\partial^2 \chi^2}{\partial \theta_i \partial \theta_j} = 
\sum_{k=1}^{N_{\rm max}}\frac{1}{\sigma (x_k)^2}\frac{\partial x_k}{\partial \theta_i}
\frac{\partial x_k}{\partial \theta_j}.
\label{eq:fisher1}
\end{equation}
The Fisher matrix is closely related to the precision with which the
parameters, $\theta_i$, can be determined.
If all free parameters are to be determined from the data alone without
any priors then it follows from the Cramer-Rao inequality
\cite{kendall} that
\begin{equation}
\sigma(\theta_i) = \sqrt{(F^{-1})_{ii}}
\end{equation}
for an optimal unbiased estimator, such as the maximum likelihood
\cite{tth}.

In order to estimate how degenerate parameter $i$ is with 
another parameter, $j$, one can calculate how $\sigma(\theta_i)$
changes if parameter $j$ is kept fixed instead of free in the
analysis. Starting from the $2 \times 2$ sub-matrix
\begin{equation}
S_{ij} = (F^{-1})_{ij},
\end{equation}
one then finds
\begin{equation}
\sigma_{j \,\, {\rm fixed}}(\theta_i) = \sqrt{\frac{1}{(S^{-1})_{ii}}}
\end{equation}

We therefore define the quantity 
\begin{equation}
r_{ij} = \frac{\sigma_{j \,\, {\rm fixed}}(\theta_i)}{\sigma (\theta_i)}
\leq 1
\label{eq:rij}
\end{equation}
as a measure of the degeneracy between parameters $i$ and $j$.

In the next two sections we discuss how to calculate the contributions
to the Fisher matrix from LSS and CMB data respectively.

\section{LSS data}

At present data from the first very large scale precision galaxy
surveys are becoming available. The 2dF survey has measured about
250,000 galaxies and the Sloan Digital Sky Survey 
\cite{sdss} is currently
in the process of measuring up to $10^6$ galaxy redshifts.

With surveys of this scale and precision it will be possible to
obtain quite precise limits on the neutrino mass. However,
even for large scale surveys the problem is that massive neutrinos
mainly affect the smaller scales where there are potential problems
with non-linearity.

At present analyses are usually carried out using an effective cut
in $k$-space at $k = 0.2 h \,\, {\rm Mpc}^{-1}$. This cut is placed
roughly where the quantity $\Delta$, defined as
\begin{equation}
\Delta^2 = \frac{V}{(2\pi)^2} 4 \pi k^3 |\delta_k|^2 \propto k^3 P(k),
\end{equation}
is equal to 1 in linear theory. Here, $P(k)$ is the power spectrum
of fluctuations.
In most cases this corresponds reasonably well to the point where
$|\Delta^2_{\rm Non-Linear}-\Delta^2_{\rm Linear}| \sim \Delta^2_{\rm Linear}$,
and the reason for the cut-off is then that poorly understood non-linear
effects begin to dominate the power spectrum at scales smaller than
the cut-off.

However, for problems involving the detection of neutrino masses in the
0.1 eV range one must be even more careful.
For instance, for $m_\nu = 0.1$ eV the power spectrum suppression
is only $\Delta P/P \sim 0.06$ so that any effect on the power spectrum
at this level which is not well understood can impair the mass detection
ability.

From simulations Peacock and Dodds \cite{peacock}
were able to derive the following
approximate relation between the linear and non-linear spectra
\begin{equation}
\Delta^2_{\rm NL} = f_{\rm NL}[\Delta^2_{\rm L}],
\end{equation}
with
\begin{equation}
f_{\rm NL}[x] = x \left( \frac{1+B \beta x + [Ax]^{\alpha \beta}}
{1+([Ax]^\alpha g^3(\Omega)/[Vx^{1/2}])^\beta}\right),
\end{equation}
and
\begin{eqnarray}
A & = & 0.482 (1+n/3)^{-0.947} \\
B & = & 0.226 (1+n/3)^{-1.778} \\
\alpha & = & 3.310 (1+n/3)^{-0.244} \\
\beta & = & 0.862 (1+n/3)^{-0.287} \\
V & = & 11.55 (1+n/3)^{-0.423} \\
g(\Omega) & = & \frac{\frac{5}{2}\Omega_m}{\Omega_m^{4/7} - \Omega_{\Lambda}
+(1+\Omega_m/2)(1+\Omega_\Lambda/70)}.
\end{eqnarray}
At the large scales we are interested in one finds that
\begin{equation}
\frac{\Delta^2_{\rm Non-Linear}-\Delta^2_{\rm Linear}}{\Delta^2_{\rm Linear}}
\to Q \Delta^2_{\rm Linear},
\end{equation}
where
\begin{equation}
Q = \frac{B \beta + A^\alpha (\alpha-\frac{1}{2}) g^3(\Omega) V^{-\beta}}
{\beta}.
\label{eq:qn}
\end{equation}
We take $\Omega_m=0.3$ and $\Omega_\Lambda=0.7$ as values for the matter
and vacuum energy densities respectively.

In the above equations $n$ is the effective spectral index of the
linear power spectrum. At very small scales $n \sim -3$ and at
scales beyond the horizon at matter-radiation equality $n \sim 1$.
Fig.~1 shows $Q(n)$. For the scales we are interested in $n \sim 0-1$
so that $Q \sim 0.2$.

\begin{figure}[t]
\begin{center}
\epsfysize=7truecm\epsfbox{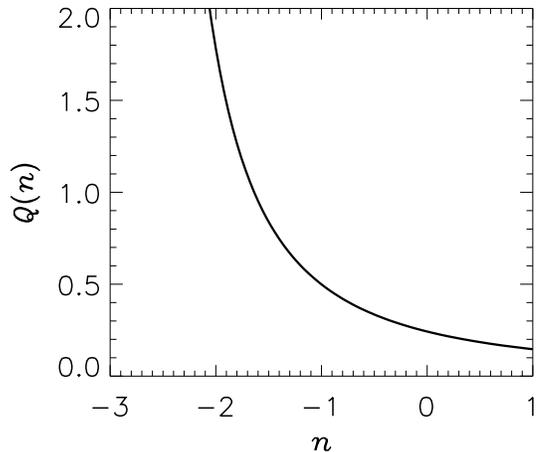}
\end{center}

\caption{The function $Q(n)$ described in Eq.~(\ref{eq:qn}).}
\label{fig1}
\end{figure}

In order to have a reasonably clean interpretation of the data a
$k$-space cut should be made roughly where
\begin{equation}
\frac{|\Delta^2_{\rm Non-Linear}-\Delta^2_{\rm Linear}|}{\Delta^2_{\rm Linear}}
\simeq \left| \frac{\Delta P(k)|_{m_\nu}}{P(k)}\right|
\end{equation}

For a neutrino mass of 0.1 eV this means that
$Q \Delta^2_{\rm Linear}|_{k_{\rm cut}} \simeq 0.06$, or that
$\Delta^2_{\rm Linear}|_{k_{\rm cut}} \simeq 0.3$. This requirement
leads to an estimate of $k_{\rm cut} \simeq 0.1 h \,\, {\rm Mpc}^{-1}$,
which in fact is not far from the 
$k_{\rm cut} = 0.2 h \,\, {\rm Mpc}^{-1}$ often used in present analyses.

\subsection{$k$-dependent bias}

In all present parameter estimation analyses it is assumed that the
bias parameter, $b^2(k) \equiv P_g(k)/P_m(k)$, is independent of the
scale, $k$.

However, many independent simulations find that bias is in fact
quite strongly scale dependent \cite{blanton,mann} in the
non-linear regime. In the linear regime bias is expected to be constant,
and the asymptotic value $b_\infty = lim_{k \to 0} b(k)$ is reached
as a scale of roughly $k \simeq 0.1-0.2  h \,\, {\rm Mpc}^{-1}$.
This means that at scales larger than $k_{\rm cut}$ bias should
be very close to scale-independent, and that we can therefore use
a single parameter, $b$, to describe it.

\subsection{Mock LSS surveys}

For purposes of parameter estimation the most important parameter
in galaxy surveys is the effective volume, defined as
\begin{equation}
V_{\rm eff} = \int \left[\frac{\bar{n}({\bf r}) P(k)}
{1+\bar{n}({\bf r}) P(k)}\right] d^3 r.
\end{equation}
In the above equation $n({\bf r})$ is the selection function. The
simple interpretation of $V$ is that it is the volume available
for measuring power at wavenumber $k$.

In the following we shall assume that the survey is volume limited,
meaning that the selection function is constant throughout the
survey volume. If the survey is flux limited the selection function
is much more complicated.
In the region where $P(k) \gtrsim 1/\bar{n}$, $V_{\rm eff}$ is
independent of $k$ and equal to the total survey volume.

Essentially this means that, with certain restrictions,
we can use just one free parameter,
$V_{\rm eff}$, to describe a hypothetical galaxy survey.

It was shown in Ref.~\cite{tegmark_lss} that the contribution
to the Fisher matrix from such a galaxy survey can be written as
\begin{equation}
F_{ij} \simeq 2 \pi \int_{k_{\rm min}}^{k_{\rm max}}
\frac{\partial \ln P(k)}{\partial \theta_i}
\frac{\partial \ln P(k)}{\partial \theta_j}
w(k) d \ln k,
\end{equation}
where the weight-function is $w(k) = V_{\rm eff}/\lambda^3$ and
$\lambda = 2\pi/k$. The upper limit of the integral should be
taken to be $k_{\rm cut}$, discussed above.
In principle the lower limit, $k_{\rm min}$, should be zero 
but at large scales the assumption that 
$P(k) \gtrsim 1/\bar{n}$ breaks down. However, by far the most
of the weight in the above integral comes from $k$ close to the
upper limit. Therefore, as long as the $k$ where 
$P(k) = 1/\bar{n}$ is much smaller than $k_{\rm max}$ the error from
taking $V_{\rm eff}$ {\it and} $k_{\rm min} = 0$ is quite small.

It should be noted that the above integral expression is quite crude.
However, it offers a very simple way of estimating parameter estimation
errors from galaxy surveys. The error arising from it can be
of order a factor 2, 
leading to an error in the estimated $\sigma (\theta_i)$
of order $2^{1/2}$.

Instead of $V_{\rm eff}$ we use 
$\lambda_{\rm eff} = (3V_{\rm eff}/4\pi)^{1/3}$ as the free parameter.
As discussed in Ref.\ \cite{tegmark_lss} the SDSS BRG survey 
\cite{sdss} has an
effective volume of roughly $(1 h^{-1} \,\, {\rm Gpc})^3$, corresponding
to $\lambda_{\rm eff} \simeq 620 h^{-1} \,\, {\rm Mpc}$.

Note that the number of independent Fourier modes on a given scale, $k$,
enclosed within the survey volume
is proportional to $V_{\rm eff}$. 
Therefore it essentially corresponds to the factor $(2l+1)$ for the
CMB measurements which measures the number of $m$-modes for a given $l$.
In that sense both $V_{\rm eff}$ and $(2l+1)$ are a measure of the
lack of ergodicity in the given data set.

\section{CMB data}

The CMB temperature
fluctuations are conveniently described in terms of the
spherical harmonics power spectrum
\begin{equation}
C_l \equiv \langle |a_{lm}|^2 \rangle,
\end{equation}
where
\begin{equation}
\frac{\Delta T}{T} (\theta,\phi) = \sum_{lm} a_{lm}Y_{lm}(\theta,\phi).
\end{equation}
Since Thomson scattering polarizes light there are additional powerspectra
coming from the polarization anisotropies. The polarization can be
divided into a curl-free $(E)$ and a curl $(B)$ component, yielding
four independent power spectra: $C_{T,l}, C_{E,l}, C_{B,l}$ and $C_{ET,l}$.
In the present work we neglect the curl-free component since it is
usually tiny compared with the other contributions. However, for some purposes,
such as the detection of tensor fluctuation modes, it is essential.
Altogether we focus on the three spectra $C_{T,l}, C_{E,l}$ and $C_{ET,l}$.

The initial measurement of CMB anisotropies by COBE \cite{Bennett:1996ce}
initiated the development of a new generation of high-precision
CMB experiments.
At present there are data sets on the temperature anisotropy from
BOOMERANG \cite{boom}, MAXIMA
\cite{max}, DASI \cite{dasi}, VSA \cite{VSA}, CBI \cite{CBI},
and several other experiments, as well as polarization data from
DASI \cite{dasi}.

In the near future data from the MAP and Planck satellites will become
available. This data offers full sky coverage combined with very
small pixel noise due to long integration times.

\subsection{Foregrounds}

There are several sources of foreground contamination in the CMB
data. Within the galaxy these are primarily synchrotron and free-free 
emission, as well as dust.
Extragalactic contamination is mainly due to point sources and 
the SZ-effect.

Several papers deal with the issue of foregrounds for future 
CMB experiments. In the present work we rely on the calculations
by Tegmark {\it et al.} \cite{foregrounds}. Their estimate is that
for the temperature anisotropies there is little effect from foregrounds
at the important frequencies around 100 GHz. The most likely source
of contamination at high $l$ is from point sources, but this effect
is unlikely to be dominant for $l \lesssim 2500$.

For the polarization anisotropy the problem with foregrounds is more
severe. For $E$-type polarization the foreground contamination
from point sources is likely to be dominant already beyond $l \sim 1500$.

In conclusion, no CMB experiment is likely to retrieve information
on the primary spectrum beyond $l \sim 2500$ for temperature and
$l \sim 1500$ for polarization. In the present work we assume that
the $E-T$ cross-correlation is subject to the same foreground 
contamination as the $E$-polarization.

Another important foreground issue is the question of secondary
CMB anisotropy generation from non-linearity, i.e.\ weak lensing and 
the Rees-Sciama effect (and of course the SZ effect discussed above).
These effects have been estimated to be extremely small for the
range of $l$-values we are interested in here (below 2500), but
can be dominant on smaller scales.

Finally, there is a possible effect from inhomogeneous reionization.
In all present CMB parameter estimation analyses reionization is
treated as a single parameter which is either the optical depth
due to reionization, $\tau$, or the redshift of reionization, $z_r$.
However, the reionization must to some extent have been inhomogeneous
and this could affect the arcminute scale CMB anisotropy.
It has been estimated for instance by Gruzinov and Hu \cite{gruzinov} that
this patchy reionization is unlikely to be important for
parameter estimation, but whether this is really the case is not yet
clear.

For lack of a better description we consider reionization in the
standard homogeneous picture using $\tau$ as a free parameter,
and note that this simplification is unlikely to cause any serious
problems \cite{gruzinov}.

\subsection{Mock CMB experiments}

In view of the discussion above we use the following prescription
for a mock CMB experiment. 
We assume it to be cosmic variance (as opposed to foreground) limited
up to some maximum $l$-value. This value can, however, be different
for temperature and polarization detection. Therefore a given hypothetical
experiment can be described by only two free parameters,
$l_{T, {\rm max}}$ and $l_{P, {\rm max}}$. For all experiments it will
be the case that $l_{T, {\rm max}} \geq l_{P, {\rm max}}$.

In this picture the MAP data will be well described by
$l_{T, {\rm max}} \simeq 1000$ and $l_{P, {\rm max}} = 0$, and
the Planck data by 
$l_{T, {\rm max}} \simeq 2500$ and $l_{P, {\rm max}} = 1500$.
In some sense Planck can therefore be regarded as the ``ultimate'' CMB
experiment because is measures all of the power spectrum parameter
space not dominated by foregrounds.

The contribution to the Fisher matrix from such a CMB experiment
is then

\begin{eqnarray}
F_{ij} & = & \sum_{l=2}^{l_{P,{\rm max}}}
\sum_{X,Y} \frac{\partial C_{l,X}}{\partial \theta_i}
{\rm Cov}^{-1} (C_{l,X},C_{l,Y})
\frac{\partial C_{l,Y}}{\partial \theta_j} \nonumber \\
&& \,\, + \sum_{l=l_{P,{\rm max}}}^{l_{T,{\rm max}}}
\frac{\partial C_{l,T}}{\partial \theta_i}
{\rm Cov}^{-1} (C_{l,T},C_{l,T})
\frac{\partial C_{l,T}}{\partial \theta_j},
\end{eqnarray}
where $X,Y = T,E,TE$.

The covariance matrices are given by \cite{wss}
\begin{eqnarray}
{\rm Cov}(C_{l,T},C_{l,T}) & = & \frac{2}{(2l+1)}C_{l,T}^2 \\
{\rm Cov}(C_{l,E},C_{l,E}) & = & \frac{2}{(2l+1)}C_{l,E}^2 \\
{\rm Cov}(C_{l,TE},C_{l,TE}) & = & \frac{2}{(2l+1)}[C_{l,T}^2
+ C_{l,T}C_{l,E}] \\
{\rm Cov}(C_{l,T},C_{l,E}) & = & \frac{2}{(2l+1)}C_{l,TE}^2 \\
{\rm Cov}(C_{l,T},C_{l,TE}) & = & \frac{2}{(2l+1)}C_{l,T}C_{l,TE} \\
{\rm Cov}(C_{l,E},C_{l,TE}) & = & \frac{2}{(2l+1)}C_{l,E}C_{l,TE}
\end{eqnarray}

It should be noted here that this approximation relies on
the assumption of $4\pi$ sky coverage and no pixel noise up to
the maximum $l$. Even though these assumptions are not realised in
any real experiment they are sufficiently accurate for estimating
the parameter estimation accuracy of a given experiment.

\section{Numerical results}

In order to calculate estimated $1\sigma$ errors on the various
cosmological parameters we need to apply the Fisher matrix analysis
to a specific cosmological model.

We choose as the reference model the generic $\Lambda$CDM model with
the following free parameters:
$\Omega_m$, the matter density, $\Omega_\Lambda$, the vacuum energy
density, $\Omega_b$, the baryon density, $H_0$, the Hubble parameter,
$n_s$, the spectral index of the primordial perturbation spectrum,
$\tau$, the optical depth to reionization, $Q$, the spectrum normalization,
$b$, the bias parameter, and $m_\nu$, the neutrino mass.
The reference model has the following parameters:
$\Omega_m = 0.3$, $\Omega_\Lambda = 0.7$, $\Omega_b h^2 = 0.02$,
$H_0 = 70 \,\, {\rm km} \, {\rm s}^{-1} \, {\rm Mpc}^{-1}$,
$n_s = 1$, $\tau=0$, and $m_\nu = 0.07$ eV.

\subsection{CMB data only}

Even though the canonical view of cosmological neutrino mass determination
is that CMB in itself can achieve little because the CMB spectrum
is insensitive to $m_\nu$, experiments such as Planck will in fact
be able to probe $m_\nu$ precisely.

As an example of the effect of massive neutrinos on the CMB we show the 
change in temperature and $E$-type polarization spectra for a
neutrino mass of 0.07 eV in Fig.~2.

\begin{figure}[t]
\begin{center}
\epsfysize=7truecm\epsfbox{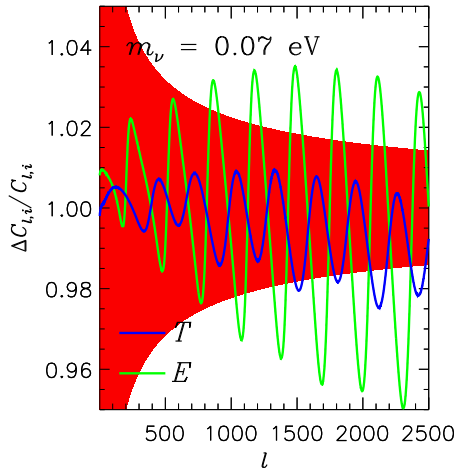}
\end{center}

\caption{The change in $C_l$ due to a neutrino mass of 0.07 eV.
The shaded band is the cosmic variance error of $\Delta C_l/C_l =
1/\sqrt{2l+1}$.}
\label{fig1}
\end{figure}

In Fig.~3 we show the expected $1\sigma$ error bar in a determination
of $m_\nu$ from a hypothetical future CMB experiment, without invoking
any other data whatsoever.

\begin{figure}[h]
\begin{center}
\epsfysize=8truecm\epsfbox{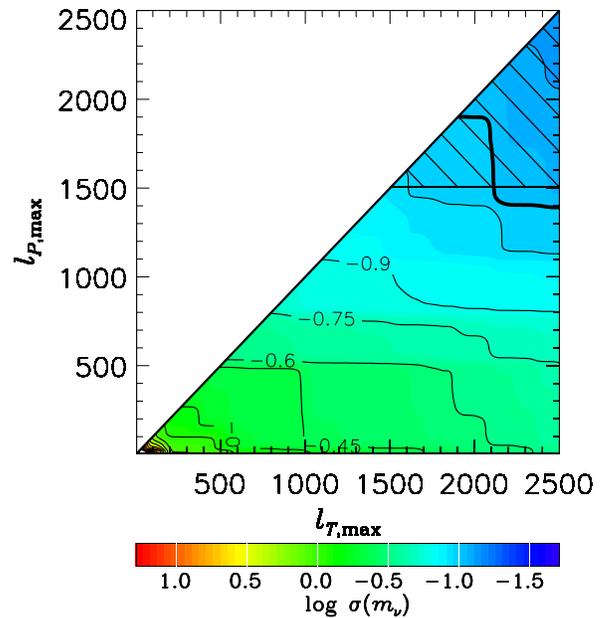}
\vspace{1truecm}
\end{center}

\caption{Estimated $1\sigma$ error bars on $m_\nu$ from the Fisher
matrix analysis using only CMB data. The thick solid line corresponds to
$m_\nu = 0.07$ eV, and the hatched region above $l_{P,{\rm max}}=1500$
is the region where the CMB signal is dominated by foregrounds.
The blank triangle in the upper left corner is the region where
$l_{P,{\rm max}} >l_{T,{\rm max}}$.}
\label{fig3}
\end{figure}

The hatched region beyond $l_{P,{\rm max}}=1500$ corresponds to the region
where the CMB signal is dominated by foregrounds, and therefore unlikely
to be retrieved in any measurement. The thick line corresponds to
$\sigma(m_\nu)=0.07$ eV, i.e.\ the central value suggested by 
current oscillation analyses provided that neutrino masses are hierarchical.

As can be seen from the figure the Planck satellite in itself without any
additional information could be able to provide a marginal detection of
a non-zero $m_\nu$. However, an unambiguous result from CMB alone
seems unlikely and in fact the argument should rather be reversed, as was
emphasized in Ref.\ \cite{Hu:1997mj}: Even a neutrino mass as small as
0.07 eV will have a significant effect on the power spectrum at high
$l$ and can bias the estimation of other parameters unless accounted
for.

\subsection{The addition of LSS data}

Adding data from LSS surveys markedly improves the ability to detect
a non-zero $m_\nu$.

In Figs.\ 4 and 5 we show how the expected $1\sigma$ error bar is
changed for the MAP and Planck experiments when LSS data is added.

\begin{figure}[h]
\begin{center}
\epsfysize=8truecm\epsfbox{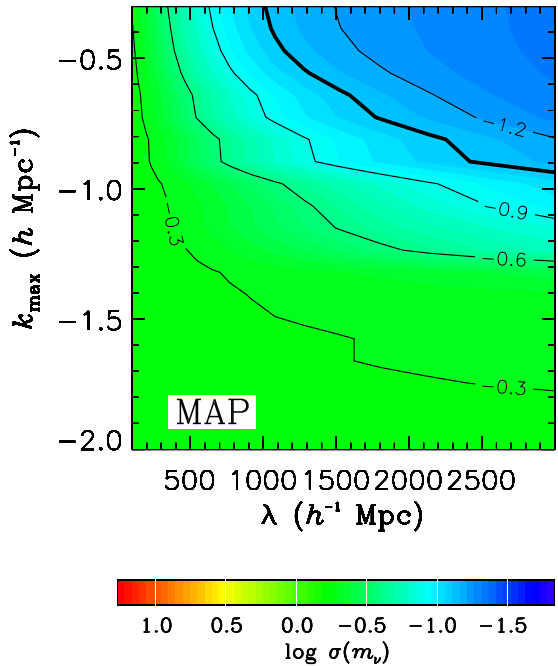}
\vspace{1truecm}
\end{center}

\caption{The estimated $1\sigma$ error bar on $m_\nu$ using CMB data
from MAP combined with a hypothetical LSS survey. The thick solid
line corresponds to
$m_\nu = 0.07$ eV.}
\label{fig4}
\end{figure}

\begin{figure}[h]
\begin{center}
\epsfysize=8truecm\epsfbox{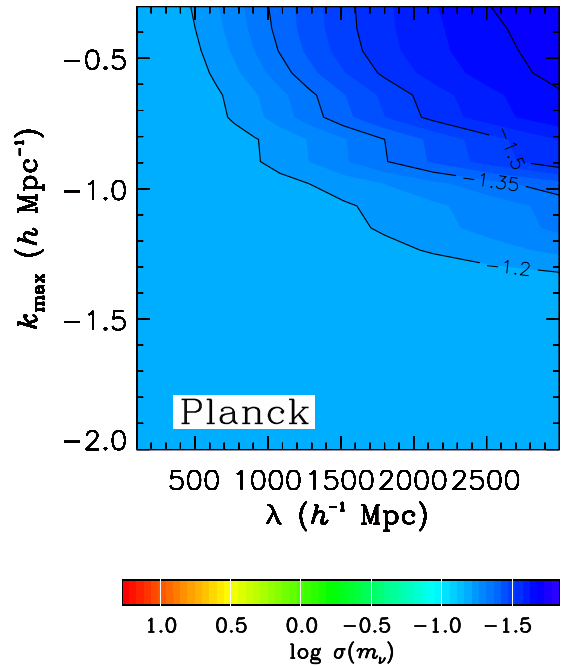}
\vspace{1truecm}
\end{center}

\caption{The estimated $1\sigma$ error bar on $m_\nu$ using CMB data
from Planck combined with a hypothetical LSS survey. The thick solid
line corresponds to
$m_\nu = 0.07$ eV.}
\label{fig5}
\end{figure}

In the near future there will be data available from the Sloan Digital
Sky Survey, for which the Bright Red Galaxy (BRG) survey in our
language corresponds to $\lambda \simeq 620 h^{-1} \,\, {\rm Mpc}$.
In Ref.~\cite{Hu:1997mj} the expected $1\sigma$ error on $m_\nu$ was
calculated for the case of combining the SDSS BRG survey and the 
MAP satellite data, taking the $k$-space cut to be at $k_{\rm cut} = 0.2
h \,\, {\rm Mpc}^{-1}$. For this specific case the error was estimated
to be $\sigma (m_\nu) \simeq 0.25-0.3$ eV. From our calculation
we find for the same case an expected error of 0.23 eV, which can
be ascribed to the fact that our estimated MAP precision is higher than
that used in Ref.~\cite{Hu:1997mj}.

For the case of Planck + SDSS we estimate the $1\sigma$ error to be
of the order 0.06 eV. This means that adding the SDSS data to the Planck
data only improves the detection threshold marginally (from
0.07 eV to 0.06 eV).

In order to achieve a $2\sigma$ detection of a neutrino mass of 0.06 eV,
which is the most favoured value using the present data, it is
necessary to combine the Planck data with data from a galaxy
survey with $\lambda \sim 3 h^{-1}$ Gpc (if $k_{\rm cut} = 0.1 h \,\,
{\rm Mpc}^{-1}$). Such a survey would contain a large fraction of the
present Hubble volume and is probably not feasible.
If, on the other hand, numerical simulations could provide the ability
to disentangle the non-linear effects beyond $k \sim 0.1 h \,\,
{\rm Mpc}^{-1}$ a much smaller survey would suffice. For instance if
$k_{\rm cut} = 0.5 h \,\, {\rm Mpc}^{-1}$ the survey volume needed
for a $2\sigma$ detection would be ``only'' 1500 $h^{-1}$ Mpc.

\subsection{Other probes}

Finally, we note that all of these calculations were done for the
case where there is no prior information on any cosmological
parameters. If the neutrino mass were the only free parameter, SDSS+Planck
could achieve a $1\sigma$ accuracy of 0.003 eV, corresponding to a
many-sigma detection of a neutrino with mass of order 0.06 eV.
This is of course wishful thinking in the sense that no other
experiment is likely to yield such information on the cosmological parameters.

In Fig.~6 we show a matrix of the parameter $r_{ij}$ for the case of 
Planck+SDSS data. We note that the main parameter degeneracies in
this case are with $H_0$ and $\Omega_b$, not with $\Omega_m$ and 
$b$ as with the present data. The estimated $1\sigma$ errors on these
two parameters from the Fisher matrix analysis are $\Delta H_0/H_0 
= 0.018$ and $\Delta \Omega_b/\Omega_b = 0.025$. If
it were possible to measure these two parameters more precisely by
other means the expected precision of the neutrino mass determination
could be improved by almost a factor 2, to a level of 0.03 eV for
SDSS+Planck.

\begin{figure}[t]
\vspace*{-0.0cm}
\begin{center}
\hspace*{-1.0cm}\epsfysize=10truecm\epsfbox{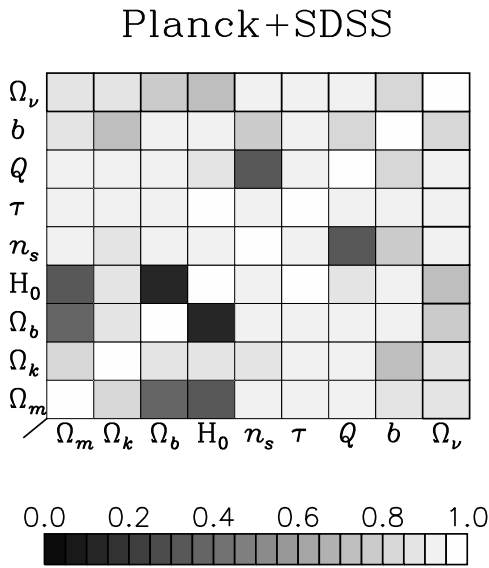}
\end{center}
\vspace*{-2cm}
\caption{Values of the parameter $r_{ij}$, defined in Eq.~(\ref{eq:rij}),
calculated for the specific case of Planck+SDSS data.}
\label{fig6}
\end{figure}

\section{discussion}

We have discussed in detail the prospects for detecting neutrino
masses with CMB and large scale structure observations, extending the
pioneering calculation in Ref.~\cite{Hu:1997mj}.

From the Sloan Digital Sky Survey and the Planck data we estimated
that it would be possible to obtain a $1\sigma$ error bar on the
neutrino mass of roughly 0.06 eV.
It should be noted that this estimate is probably fairly safe in the
sense that it requires only robust assumptions. For instance it
requires only LSS data at scales larger than 
$k \simeq 0.1 h \,\, {\rm Mpc}^{-1}$ where effects of non-linearity are
negligible.

This data would allow for a marginal detection of the neutrino mass
if the neutrino mass is 0.06 eV, the value favoured by current
oscillation data. Maltoni {\it et al.} \cite{Maltoni:2002ni} find
that the most massive mass eigenstate should be at $m \sim 0.04-0.07$ eV
(assuming the LMA Solar solution). 
Therefore it seems unlikely that neutrino masses will be detected by
SDSS+Planck data, {\it unless} additional information can be retrieved.
Perhaps the most promising prospect for this is that detailed numerical
simulations will allow for a disentanglement of the non-linear effects
i LSS data at high $k$. If SDSS data up to
$k \simeq 0.5 h \,\, {\rm Mpc}^{-1}$ could be used the error would 
diminish to 0.04 eV.

However, in order to achieve an unambiguous detection of a neutrino
mass of 0.06 eV a much larger survey than the SDSS is needed. Whether
such a survey becomes feasible in the future remains to be seen, but with
improvement in detector technology and the easier access to 8 meter 
class telescopes it may not be impossible.

It should also be noted that in the case of an inverted mass hierarchy
there will be two mass eigenstates of almost degenerate mass, so that
the atmospheric neutrino data indicates a sum of order 0.12 eV.
Such a scenario would be detectable at the $2\sigma$ level with Planck+SDSS.

All of the mass constraints discussed here should be compared with
the expected precision of other, planned experiments. The 
KATRIN tritium endpoint experiment \cite{katrin} is planned to measure
the effective electron neutrino mass, 
$m_{\nu_e} = \left( \sum_i  |U_{ei}|^2 m_i^2\right)^{1/2}$,
to an accuracy of 0.35 eV (95\% conf.). This is better than the
95\% accuracy expected from MAP+SDSS (0.45 eV at 95 \% conf.), 
but not competitive with the 0.12 eV (95\% conf.) expected from SDSS+Planck.
However, it is possible that tritium decay experiments can be pushed
to the limit of $m_{\nu_e} \lesssim 0.06$ eV, which would indicate 
hierarchical masses.

If neutrinos are Majorana particles processes such as neutrinoless
double beta decay will become possible. The detection of $2\beta0\nu$
decays probes the mass combination
\begin{equation}
m_{ee} = \left| \sum_j U^2_{ej} m_{\nu_j} \right| ,
\end{equation}
and has led to a current upper bound of 0.27 eV \cite{klapdor}.
Future experiments, such as GENIUS \cite{genius}, 
could take the accuracy to 0.01 eV,
much better than the expected accuracy from cosmology. However, this
requires that neutrinos are Majorana particles and that there
is no cancellation of terms in the expression for $m_{ee}$.

In conclusion, using LSS and CMB
data which will become available within the next 
8-10 years and reasonable assumptions it should become possible to
achieve a 95\% confidence limit on the mass of the heaviest neutrino
mass eigenstate of 0.12 eV. This is almost at the level expected if
neutrino masses are hierarchical, and with a future generation of
LSS survey it seems feasible to detect neutrino masses as low as 
0.03-0.06 eV at 95\% confidence. The timescale for such future surveys
is comparable to, or shorter than, the timescale for building tritium
endpoint experiments beyond KATRIN.
Therefore there is good reason to be optimistic regarding the 
future potential of cosmology to probe neutrino masses.

\end{document}